\documentclass[12pt]{article}
\usepackage{latexsym}

\def\tr{{\rm tr}}

\def\ket#1{\mid~\!\!\!{#1}~\!\!\rangle}
\def\bra#1{\langle~\!\!{#1}~\!\!\!\mid}

\def\QM{quantum mechanics }
\def\qm{quantum mechanics}
\def\QMl{quantum-mechanical }
\def\qml{quantum-mechanical}

\def\ON{orthonormal }
\def\${\enskip$}
\def\M{measurement }
\def\m{measurement}
\def\Q{quantum }

\begin{document}

{\bf \large \noindent On EPR-type
Entanglement in the Experiments of
Scully et Al. I. The Micromaser Case\\
and Delayed-Choice Quantum Erasure}\\

{\bf \noindent F. Herbut}\\

\vspace{0.5cm}

\rm

\noindent {\bf Abstract} Delayed-choice
erasure is investigated in two-photon
two-slit experiments that are
generalizations of the micromaser
experiment of Scully et al. [Scully, M.
O. et al. Nature {\bf 351}, 111-116
(1991)]. Applying \QM to the
localization detector, it is shown that
erasure with delayed choice in the
sense of Scully, has an analogous
structure as simple erasure. The
description goes beyond probabilities.
The EPR-type disentanglement,
consisting in two mutually incompatible
distant measurements, is used as a
general framework in both parts of this
study. Two simple coherence cases are
shown to emerge naturally, and they are
precisely the two experiments of Scully
et al. The treatment seems to require
the
relative-reality-of-unitarily-evolving-state
(RRUES) approach
. Besides insight in
the experiments, this study has also
the goal of insight in \qm . The
question is if the latter can be more
than just a "book-keeping device" for
calculating probabilities as Scully et
al. modestly
and cautiously claim.\\

\noindent {\bf Keywords} Delayed-choice
erasure. Distant measurement. EPR-type
disentanglement. Detector as \QMl
system.
Relative-reality-of-unitarily-evolving-state
approach. Relative-state
interpretation.\\

\vspace{1.5cm}

{\bf \noindent 1 Introduction}\\

\noindent The fascinating process of
quantum erasure (of entanglement), even
its more sophisticated delayed-choice
version, were first discussed by Edwin
Jaynes \cite{Jaynes}, but, as far as I
can tell, without giving any terms for
the general phenomenon.

{\footnotesize \rm \noindent
\rule[0mm]{4.62cm}{0.5mm}

\noindent F. Herbut (mail)\\
Serbian Academy of Sciences and Arts,
Knez Mihajlova 35, 11000 Belgrade,
Serbia\\
e-mail: fedorh@infosky.net and from USA
etc. fedorh@mi.sanu.ac.yu}

\noindent  Marlan O. Scully and
coworkers elaborated it [2],
\cite{Nature}.

Particularly baffling is the
delayed-choice (or after-detection)
version of erasure in contrast to the
simple(or before-detection) case. (One
should distinguish the mentioned
delayed choice in the sense of Scully
from that in the sense of Wheeler
\cite{Wheeler}.)

One of the first attempts to perform a
real erasure experiment \cite{Kwiat}
presented its theoretical part in
second quantization. However, it was
demonstrated \cite{FHMV}, along the
same lines as in this study, that
first-quantization \QMl insight is
feasible and useful. But, for
consistency, this line of approach
seems to require the {\it
relative-reality-of-unitarily-evolving-state}
or, shortly the RRUES, interpretation,
which will be explained in this
article.

The claim of Scully and Walther
\cite{ScullyFP98} that there is no
essential difference between simple and
delayed-choice erasures is further
elaborated in this article using the
reality-of-states approach. The slight
difference that does exist will
also be discussed.\\

One should clarify, for the reader's
benefit, that a bipartite state is {\it
correlated} whenever it is not
factorizable (tensorically) into the
states of its subsystems. Not all
correlations are entanglement.
Separable mixed states, i. e., ones
that are mixtures of uncorrelated
states, are correlated (unless they
have only one term),  but have no
entanglement. Contrariwise, the
correlations in any bipartite pure
state consist only of entanglement.

The notion of an EPR-type bipartite
pure state, or, equivalently one that
contains {\it EPR-type entanglement},
also requires clarification. By
definition, \$\ket{\Phi}_{I,II}\$ is an
EPR-type bipartite state vector if one
can have {\it EPR-type
disentanglement}, i. e., if one can
perform {\it distant}
(direct-interaction-free) {\it \M } (cf
\cite{FHdistmeas}) {\it of either of
two mutually incompatible subsystem
observables}.

The notion of {\it disentanglement} (as
well as of entanglement) goes back to
Schr\"{o}dinger \cite{Schr}.\\

\vspace{0.5cm}

{\bf \noindent 2 Does Interaction
Destroy Coherence?}\\

\noindent The authors remark in their
thorough analysis of erasure
\cite{ScullyFP98} (p. 399):

\begin{quote} \small "Just how the
acquisition of which way ({\it Welcher
Weg}) information rubs out the
interference fringe is an interesting
question."
\end{quote}

\rm

As a contrast to their answer, they
quote Feynman, who comments on Young's
two-slit interference case
\cite{Young}, \cite{Feynman}:

\begin{quote} \small "If an apparatus is
capable of determining which hole the
electron (or atom or ...) goes through,
it cannot be so delicate that it does
not disturb the pattern in an essential
way."
\end{quote}

\rm

The answer of the authors then goes as
follows (\cite{ScullyFP98}, p. 400).

\begin{quote}
\small "... it is not necessarily the
indelicate nature of our probing that
rubs out the interference pattern. It
is simply {\it knowing} (or having the
ability to know even if we choose not
to look at the {\it Welcher Weg}
detector) which eliminates the pattern.
This has been verified experimentally."
\end{quote}

\rm

I agree, but I'd like to approach the
matter differently to gain additional
insight.

Let us take a simple case of coherence
like e. g.
$$\ket{\phi}_{II}\equiv \alpha\ket{1}_{II}+\beta
\ket{2}_{II},\eqno{(1a)}$$
$$\bra{i}_{II}\ket{j}_{II}=\delta_{ij},\quad
i,j=1,2;\quad |\alpha|^2+|\beta|^2=1,
\quad \alpha\not=
0\not=\beta.\eqno{(1b)}$$

The state vectors
\$\ket{j}_{II},\enskip j=1,2,\$ may, e.
g., describe transition of the slits in
the Young experiment. Then
\$\ket{\phi}_{II}\$ of (1a) is the
coherent state that gives fringes on
the screen with detectors.

If another system \$I\$ has been
interacting with the system described
by \$\ket{\phi}_{II}\$ in the way of
ideal measurement, or some similar
process has taken place, which has led
to a bipartite state vector
$$\ket{\Phi}_{I,II}=\alpha'\ket{1}_I
\ket{1}_{II}+\beta'\ket{2}_I\ket{2}_{II},
\quad |\alpha'|^2+|\beta'|^2=1, \quad
\alpha'\not= 0\not=\beta' \eqno{(2)}$$
with \$\bra{i}_I\ket{j}_I
=\delta_{ij},\enskip i,j=1,2,\$ then
the subsystem state (reduced density
operator)
\$\rho_{II}\equiv\tr_I\Big(
\ket{\Phi}_{I,II}\bra{\Phi}_{I,II}\Big)\$
of subsystem \$II\$ is
$$\rho_{II}=|\alpha'|^2\ket{1}_{II}
\bra{1}_{II} +|\beta'|^2
\ket{2}_{II}\bra{2}_{II}.\eqno{(3)}$$

Comparing (1a) and (3), we see that
coherence has been eliminated in
subsystem \$II,\$ but it is {\it not
destroyed}. We see in (2) that
coherence reappears in the state of the
composite system. {\it Coherence being
elevated to the larger system by
interaction of the parts} or by some
analogous process is a remarkable fact
in \QM (cf \cite{Zeh}). It is an
obvious consequence of the linear
nature of the unitary evolution
operator of the composite system (which
includes the interaction etc.).

{\it Coherence} in
\$\ket{\Phi}_{I,II}\$ of (2) {\it
implies entanglement} between the
subsystems.\\

\vspace{0.5cm}

{\bf \noindent 3 Simplest EPR-type
Entanglement}\\

\noindent Let us assume that a
bipartite state vector \$\ket{\Psi
}_{I,II}\$ is given. It can {\it
always} be written in the form of a
so-called {\it canonical Schmidt
decomposition}, which is in terms of
bi-orthonormal bases with positive
expansion coefficients. (For a concise
review see Subsection 2.1 in
\cite{envariance}.)

To treat the simplest case, we assume
that the so-called Schmidt rank, i. e.,
the number of terms in the
decomposition (which is an invariant
with respect to the different
biorthogonal decompositions) is two:
$$\ket{\Psi }_{I,II}=r_1^{1/2}\ket{1}_I
\ket{1}_{II}+r_2^{1/2}\ket{2}_I\ket{2}_{II}.
\eqno{(4)}$$

A practical advantage of the canonical
Schmidt decomposition is that one can
read in it the spectral forms of the
subsystem states (reduced density
operators):
$$\rho_I\equiv \tr_{II}\Big(\ket{\Psi
}_{I,II}\bra{\Psi
}_{I,II}\Big)=r_1\ket{1}_I\bra{1}_I+
r_2\ket{2}_I\bra{2}_I,\eqno{(5a)}$$
$$\rho_{II}=r_1\ket{1}_{II}\bra{1}_{II}+
r_2\ket{2}_{II}\bra{2}_{II}.\eqno{(5b)}$$

{\it Physically}, a canonical Schmidt
decomposition, like (4), plays an
important role in {\it distant \m }. If
one measures (in a predictive way) the
observable consisting in determining if
the subsystem \$I\$ is in the state
\$\ket{1}_I\$ or \$\ket{2}_I\$, then
one has the change of state from (4) to
\$\ket{j}_I\ket{j}_{II},\enskip
j=1\enskip\mbox{or}\enskip 2\$, which
implies that the (technically)
'distant' subsystem \$II\$ is brought
into the state \$\ket{j}_{II},\enskip
j=1,2\$. ('Distantness' consists in
lack of interaction between subsystems
\$I\$ and \$II\$.) Distant \M takes
place without interaction of the
measuring apparatus with the distantly
measured subsystem (\$II\$ in our
example). It is a {\it consequence} of
the direct \M on subsystem \$I\$ and
{\it of the quantum correlations}
between subsystems \$I\$ and \$II$.

The very possibility of the described
distant \M is {\it the which-way
knowledge} that subsystem \$I\$ has
about subsystem \$II\$ (and {\it vice
versa}). If we do perform the \m, then
also we acquire this knowledge.

In general, one has {\it EPR-type
entanglement} if there is degeneracy in
at least one of the positive
eigenvalues of the reduced density
operators \cite{FHEPR}. (For EPR-type
entanglement see also
\cite{ScullyEPR}). In (4)
 this means
\$r_1=r_2.\$ Hence, (4) and (5b) become
$$\ket{\Psi}_{I,II}=(1/2)^{1/2}\Big(
\ket{1}_I
\ket{1}_{II}+\ket{2}_I\ket{2}_{II}
\Big). \eqno{(6a)}$$ and
$$\rho_{II}=(1/2)\ket{1}_{II}
\bra{1}_{II}+(1/2)\ket{2}_{II}\bra{2}_{II}
\eqno{(6b)}$$ respectively.

Now the canonical Schmidt decomposition
has a non-denumerably infinite
degeneracy, viz., every \ON basis in
the two-dimensional range of \$\rho_I\$
is part of an an eigen-basis of
\$\rho_I\$ (and symmetrically for
\$\rho_{II}\$), and (generalized)
expansion in it leads to a canonical
Schmidt decomposition (cf Subsection
2.1 in \cite{envariance}).

To be concrete, let $$\ket{a}_I\equiv
e^{i\lambda}p\ket{1}_I+e^{i\delta}q
\ket{2}_I,\quad 0<p,q<1,\quad
p^2+q^2=1,\quad 0\leq \lambda, \delta
<2\pi\eqno{(7a)}$$ be an arbitrary {\it
coherence state vector} (a linear
combination with both coefficients
non-zero) in the range of \$\rho_I\$.
Then, as easily checked, the obviously
unique (up to a phase factor) {\it
orthogonal state vector} has the form
$$\ket{b}_I\equiv e^{i\gamma}q
\ket{1}_I +e^{i(\gamma +\delta-\lambda
+\pi)}p\ket{2}_I,\quad 0\leq\gamma
<2\pi.\eqno{(7b)}$$

To expand the state vector
\$\ket{\Psi}_{I,II}\$ (cf (6a)) in the
sub-basis \$\{\ket{a}_I\$,
\$\ket{b}_I\}\$, we define
$$\overline{\ket{a}}_{II}\equiv
e^{-i\lambda}p\ket{1}_{II}+e^{-i\delta}q
\ket{2}_{II},\qquad
\overline{\ket{b}}_{II}\equiv
e^{-i\gamma}q \ket{1}_{II}
+e^{-i(\gamma +\delta-\lambda
+\pi)}p\ket{2}_{II}. \eqno{(8a,b)}$$

Then, as shown in Appendix A, the same
bipartite state vector (6a) has the
alternative canonical Schmidt
decomposition
$$\ket{\Psi}_{I,II}=(1/2)^{1/2}
\Big(\ket{a}_I\overline{\ket{a}}_{II}+
\ket{b}_I\overline{\ket{b}}_{II}\Big).
\eqno{(9)}$$

The physical meaning of (9) is
analogous to that of (6a) (or (4))
explained above: If one performs a
direct \M on subsystem \$I\$ to find
out if it is in the state \$\ket{a}_I\$
or \$\ket{b}_I\$, {\it ipso facto} one
finds subsystem \$II\$ (by distant \m )
in the corresponding, i. e., 'partner',
state \$\overline{\ket{a}}_{II}\$ or
\$\overline{\ket{b}}_{II}\$
respectively.

The possibility to perform, in
principle, either of the two {\it
mutually incompatible} distant \m s
based on (6a) or (9) respectively, is
called {\it EPR-type disentanglement}.
(The original EPR paper \cite{EPR}
dealt with EPR-type disentanglement of
position and linear momentum.) A case
of EPR-type disentanglement performed
in {\it one} real experiment will be
discussed in Part II of this study.

Distant \M based on the basic canonical
Schmidt decomposition (6a) is often
feasible. In all versions of the Young
two-slit experiment with pairs of
particles it is called 'which-way' \m .

Concerning {\it linearly polarized
photons}, when in \$\ket{j}\$ \$j\$
refers to two mutually orthogonal
linear polarizations, experiments that
establish the EPR-type entanglement in
(6a) have been performed. (See the
\$J=0\enskip\rightarrow\enskip
J=1\enskip\rightarrow\enskip J=0\$
atomic cascade transitions as source of
photon pairs -\$J\$ being the atomic
angular momentum \cite{Aspect}.)

If in \$\ket{j}\$ \$j\$ in (6a) means
spin-up and spin-down of a
spin-one-half particle, then (6a) gives
one of the vectors in the so-called
Bell operator basis used in the
discovery of teleportation
\cite{TELEP}. Distant spin-projection
\M is here certainly feasible.

Thus, EPR-type correlations have been
studied in great detail theoretically
and experimentally, but, as far as I
know, without actual EPR-type
disentanglement experiments.

Performing distant \M based on the
canonical Schmidt decomposition (9)
requires the possibility to measure the
basis (7a,b) in direct \m . This takes
exceptional experimental skill.

It will be shown that the experiments
of Scully et al. that appeared in the
literature are restricted to {\it
simple coherence bases}, the first one:
$$\ket{\pm }_I\equiv (1/2)^{1/2}\Big(
\ket{1}_I\pm\ket{2}_I\Big),\eqno{(10)}$$
and the second one
$$\ket{\pm i}_I\equiv
(1/2)^{1/2}\Big( \ket{1}_I\pm
 i\ket{2}_I\Big).\eqno{(11)}$$ (It is
 obvious from (7a) and (7b) that (10)
 and also (11) are orthogonal sets.)

These bases are not just mathematically
simple. They have simple physical
properties with respect to
\$\ket{\Psi}_{I,II}\$ (cf (6a)).
Namely, the {\it first} simple
coherence basis \$\ket{\pm }_I\$, which
is given by (10), is the only coherence
basis determining, via expansion of
\$\ket{\Psi }_{I,II}\$, a canonical
Schmidt decomposition in which each
term is symmetric under the
(two-particle) exchange operator
(assuming \$p=q\$ in (7a) for
simplicity):
$$\ket{\Psi}_{I,II}=(1/2)^{1/2}\bigg\{
\Big[(1/2)^{1/2}(\ket{1}_I+\ket{2}_I)\Big]
\Big[(1/2)^{1/2}(\ket{1}_{II}+\ket{2}_{II})\Big]
\quad +$$
$$\Big[(1/2)^{1/2}
(\ket{1}_I-\ket{2}_I)\Big]\Big[(1/2)^{1/2}
(\ket{1}_{II}-\ket{2}_{II})\Big]\bigg\}
\eqno{(12)}$$ (cf (9) with (8a,b)).
This property is {\it physical due to
the distant-\M interpretation of the
decomposition} (explained above).

The {\it second} simple coherence basis
\$\ket{\pm i}_I\$ (cf (11)) is
determined as being the only coherence
basis in the corresponding canonical
Schmidt decomposition of which the
exchange operator maps the first term
on the rhs onto the second one and {\it
vice versa}:
$$\ket{\Psi}_{I,II}=(1/2)^{1/2}\bigg\{
\Big[(1/2)^{1/2}(\ket{1}_I+i\ket{2}_I)\Big]
\Big[(1/2)^{1/2}(\ket{1}_{II}-i\ket{2}_{II})
\Big]\quad +$$
$$\Big[(1/2)^{1/2}
(\ket{1}_I-i\ket{2}_I)\Big]\Big[(1/2)^{1/2}
(\ket{1}_{II}+i\ket{2}_{II})\Big]\bigg\}.
\eqno{(13)}$$ These claims are proved
in Appendix B.

The coherence canonical Schmidt
decompositions (12) and (13) obviously
imply the following decompositions of
the state of subsystem \$II\$, which
consists of coherence terms:
$$\rho_{II}=(1/2)\bigg\{
\Big[(1/2)^{1/2}(\ket{1}_{II}+\ket{2}_{II})\Big]
\Big[(1/2)^{1/2}(\bra{1}_{II}+\bra{2}_{II})\Big]
\quad +$$ $$\Big[(1/2)^{1/2}
(\ket{1}_{II}-\ket{2}_{II})\Big]\Big[(1/2)^{1/2}
(\bra{1}_{II}-\bra{2}_{II})\Big]\bigg\},
\eqno{(14)}$$ and
$$\rho_{II}=(1/2)\bigg\{
\Big[(1/2)^{1/2}(\ket{1}_{II}+i\ket{2}_{II})\Big]
\Big[(1/2)^{1/2}(\bra{1}_{II}-i\bra{2}_{II})\Big]
\quad +$$ $$\Big[(1/2)^{1/2}
(\ket{1}_{II}-i\ket{2}_{II})\Big]\Big[(1/2)^{1/2}
(\bra{1}_{II}+i\bra{2}_{II})\Big]\bigg\}.
\eqno{(15)}$$\\ (Note that also (6b) is
valid).

The well-known micromaser thought
experiment of Scully et al.
\cite{Nature} was based on the simple
coherence basis (10) and the
corresponding canonical Schmidt
decomposition (12). In outline the
experiment goes as follows.

A plane atom wave split into two
coherent collimated beams enters
microwave cavities. Its passage through
the cavities forces the previously
excited atom to emit a photon
(subsystem \$I\$). As long as the
shutters remain closed, the photon is
found inside one of the cavities, and
which-way information is obtained, i.
e., \$\ket{1}_I\bra{1}_I\$ or
\$\ket{2}_I\bra{2}_I\$ is detected. No
interference is observed on the atoms
(subsystem \$II\$, cf the two terms on
the rhs of (6b)).

When the shutters are opened, the
which-way information is erased: Atoms
associated with the photons that are
subsequently detected by a photo-sensor
give rise to the same interference
pattern that would be observed without
the cavities (cf (1a) with \$\alpha
=\beta =(1/2)^{1/2}\$) because it is
the symmetric field \$(1/2)^{1/2}\Big(
\ket{1}_I+\ket{2}_I\Big)\$ that is
actually detected. The remaining atoms
(subsystems \$II\$ in the state
\$(1/2)^{1/2}\Big(
\ket{1}_{II}-\ket{2}_{II}\Big)\$)
correspond to the antisymmetric field,
and they give rise to the complementary
interference pattern.

This is a beautiful illustration of
EPR-type disentanglement because both
mutually incompatible distant
measurements are seen to be performable
(at least in principle).

The remarkable real random-choice and
delayed-choice erasure experiment of
Kim et al. \cite{KIM} was based on the
second simple coherence basis (11) and
the corresponding canonical Schmidt
decomposition (13). This will be proved
in the second part of this study.\\

It should be noted that {\it coherence
is a relative notion}: The state
vectors \$\ket{+}_{II}\$ and
\$\ket{-}_{II}\$ are coherent with
respect to the state vectors
\$\ket{1}_{II}\$ and \$\ket{2}_{II},\$
which are assumed to have basic
physical meaning in the experiment
discussed (e. g., passing the slits
\$1\$ and \$2\$ respectively).
Mathematically, also the reverse
relation is true: the latter two state
vectors are coherent with respect to
the former two. But this is physically
irrelevant in the case at issue.\\

If the interaction described in the
preceding section has rubbed out
interference, then the coherence
distant \M based on (10) and (12)
reestablishes coherence in subsystem
\$II\$. This is {\it erasure} (of
entanglement). If direct \M of the
states \$\ket{\pm }_I\$ is performed
{\it before or after localization} \M
in the state \$\rho_{II}\$,  we have
{\it simple erasure} (before-detection
erasure) or {\it delayed-choice
erasure} (after-detection erasure)
respectively.

It will be shown that both types of
erasure are EPR-type disentanglement
with the only difference that in simple
erasure no detector has to be included
in the quantum-mechanical description,
whereas in delayed-choice erasure, to
'see' the EPR-type disentanglement, \QM
has to be applied also to the
localization detectors for particle
\$II\$.\\

We are now going to prepare the ground
for treatment of delayed-choice
erasure.\\

\vspace{0.5cm}

{\bf \noindent 4 Subsystem
Interaction}\\

\noindent Let us assume that subsystem
\$II\$ interacts with a third system
\$D_{II}\$ (we have primarily a
detector in mind) in some state that we
denote by \$\ket{0}_{D_{II}}\$ while
the former is a subsystem of the
composite system in the state
\$\ket{\Psi}_{I,II}\$ (cf (6a) or (9)).
Let us describe the interaction by a
{\it unitary evolution operator}
\$U_{IID_{II}}.\$ Also evolution for
some time after the interaction has
ceased may be included. For simplicity,
we disregard the possible changes
caused by  evolution of subsystem \$I$.

The complementary (which-way and
coherence) decompositions (6a) and (9)
lead to
$$\Big(1_I\otimes
U_{IID_{II}}\Big)\Big(\ket{\Psi
}_{I,II}\ket{0}_{D_{II}}\Big)=$$
$$(1/2)^{1/2}\ket{1}_I\otimes
\Big[U_{IID_{II}}\Big(
\ket{1}_{II}\ket{0}_{D_{II}}\Big)\Big]
\enskip+$$
$$(1/2)^{1/2}
\ket{2}_I\otimes\Big[
U_{IID_{II}}\Big(\ket{2}_{II}
\ket{0}_{D_{II}}
\Big)\Big],\eqno{(16)}$$ and
$$\Big(1_I\otimes U_{IID_{II}}\Big)\Big(
\ket{\Psi
}_{I,II}\ket{0}_{D_{II}}\Big)=$$
$$(1/2)^{1/2}\ket{a}_I\otimes
\Big[U_{IID_{II}}\Big(\overline{
\ket{a}}_{II}\ket{0}_{D_{II}}\Big)\Big]
\enskip +$$
$$(1/2)^{1/2}
\ket{b}_I\otimes
\Big[U_{IID_{II}}\Big(\overline{\ket{b}}
_{II}\ket{0}_{D_{II}}\Big)\Big]
\eqno{(17)}$$ respectively.

Since \$\ket{1}_{II}\$ and
\$\ket{2}_{II}\$ are orthogonal, so are
\$\Big(\ket{1}_{II}
\ket{0}_{D_{II}}\Big)\$ and
\$\Big(\ket{2}_b\ket{0}_{D_{II}}\Big),\$
and, on account of the unitary nature
of the evolution operator, so are also
\$U_{IID_{II}}\Big(
\ket{1}_{II}\ket{0}_{D_{II}}\Big)\$ and
\$U_{IID_{II}}\Big(\ket{2}_{II}
\ket{0}_{D_{II}}\Big).\$ The analogous
argument holds for evolution (17).
Hence, (16) and (17) are still two
complementary canonical Schmidt
decompositions of a bipartite
\$I+IID_{II}\$ {\it EPR-type state
vector}. Viewing the lhs of these
relations as a tripartite state,
clearly the state of the subsystem
\$I+II\$ does no longer have the nice
EPR-type coherence of (6a) or (9). (The
change depends on the structure of the
evolution operator).

To simplify the exposition, we now
confine the general coherence basis
\$\{\ket{a}_I,\enskip \ket{b}_I\}\$ (cf
(7a) and (7b)) to the two simple
choices (cf (10) and (11)), which are
the only relevant ones in this study.
(However, the argument is valid for the
general case.)

Further, one can make use of the
linearity of the evolution operator.
Then substitution of (10) in (17) gives
$$U_{IID_{II}}\Big(\overline{\ket{\pm }}
_{II}\ket{0}_{D_{II}}\Big)=(1/2)^{1/2}
U_{IID_{II}}\Big(
\ket{1}_{II}\ket{0}_{D_{II}}\Big)
\enskip\pm$$
$$(1/2)^{1/2}U_{IID_{II}}
\Big(
\ket{2}_{II}\ket{0}_{D_{II}}\Big).
\eqno{(18)}$$ An analogous argument is
valid for substitution of the second
simple coherence basis \$\ket{\pm i}\$
(cf (11)) in (17).

Taking the concrete case of Young's
two-slit interference in some version,
one can view
\$U_{IID_{II}}\Big(\ket{j}_{II}
\ket{0}_{D_{II}}\Big), \enskip j=1,2,\$
as describing the sub-ensemble of atoms
(photons etc.) that have passed slit
\$j\$ and have interacted afterwards.
Then these are still the definite-way
states without coherence. Contrariwise,
the states
\$U_{IID_{II}}\Big(\overline{\ket{\pm
}}_{II}\ket{0}_{D_{II}}\Big)\$, and the
states
\$U_{IID_{II}}\Big(\overline{\ket{\pm
i}}_{II}\ket{0}_{D_{II}}\Big)\$ are
coherence states as clear from (18).

It is very important to note that
system \$D_{II}\$ is, in principle,
{\it any system} (that can interact
with subsystem \$II.\$) Further, we are
dealing with {\it any interaction}.

One should also note that one has a
composite EPR-type state vector of the
bipartite system \$D_II+IID_{II}\$ also
if any system \$D_I\$ interacts in any
way with subsystem \$I$. Namely, any
unitary evolution operator \$U_{D_{I}I}
\otimes U_{IID_{II}}\$ does not change
the form of the canonical Schmidt
dacomposition of the composite state as
long as there is no inteaction between
subsystems \$D_II\$ and \$IID_{II}\$.
(For more details, see Appendix C.)\\

\pagebreak

{\bf \noindent 5 Delayed-choice Erasure}\\

\noindent Now we make the often done,
but controversial, {\it assumption that
also macroscopic systems}, in
particular detectors, which are usually
described by classical physics, can, in
principle, be treated by the \QMl
formalism, and that they can be in pure
states.

Let system \$D_{II}\$ in the preceding
section be the detector (or system of
detectors) measuring the localization
of particle \$II\$. Then, as it was
argued above, the states
\$U_{IID_{II}}\Big(\overline{\ket{\pm
}}_{II}\ket{0}_{D_{II}}\Big)\$ or
\$U_{IID_{II}}\Big(\overline{\ket{\pm
i}}_{II}\ket{0}_{D_{II}}\Big)\$ (cf
(10) and (11)) are obtained in
delayed-choice erasure, i. e., if the
detection takes place before the direct
\M on subsystem \$I\$ ascertaining if
it is in the state \$\ket{+}_I\$ or
\$\ket{-}_I\$ is performed (analogously
for the second simple coherence
basis).\\

It was demonstrated \cite{ScullyAJP99}
(Subsection III.A) that, as far as the
relevant probabilities are concerned,
there is no difference between simple
and delayed-choice erasure. The
straightforward demonstration of the
authors is directly applicable to the
more general case treated in this
article because it is due, as they
point out, to the mathematical
properties of the tensor product.

To enable the reader to obtain a fuller
appreciation of the approach of this
paper, a treatment of delayed-choice
erasure that implies the claimed
equality of the relevant probabilities
is presented in Appendix D.

The EPR-type entanglement discussed in
this article is {\it mutatis mutandis}
applicable also to the mentioned real
experiment reported by Kim, Yu, Kulik,
Shih, and Scully \cite{KIM}. It
contains an additional upgrading of
delayed-choice erasure: it is no longer
the experimenter, but an automatic
random choice that determines if
which-way or coherence distant \M is
taking place. It can also be viewed as
an EPR-type disentanglement in terms of
two mutually incompatible distant \m s.
This is explained in the second part of
the present study.\\

\vspace{0.5cm}

{\bf \noindent 6 Summing Up}\\

\noindent We now sum up the conceptual
steps involved in {\it delayed-choice
erasure viewed as EPR-type
disentanglement}.

(i) Some mechanism rubs out the {\it
coherence} in system \$II\$ (cf (1a))
{\it elevating it} it to the larger
system \$I+II\$ (cf (2)). This implies
entanglement between the subsystems
enabling e. g. subsystem \$I\$ to
'mark' in terms of its states
\$\ket{q}_I,\enskip q=1,2,\$ the
corresponding states of subsystem
\$II\$ (cf (2)). This is a kind of
'knowledge' on part of subsystem \$I\$
about subsystem \$II$.

(ii) Performing direct \M on subsystem
\$I\$ in the bipartite state (6a) to
learn in which of the states
\$\ket{j}_I,\enskip j=1,2,\$ it is
found, one, by this very act, performs
{\it distant \M } on subsystem \$II\$
finding out if it is in the state
\$\ket{1}_{II}\$ or \$\ket{2}_{II}.\$
(For a more general discussion of
distant \M see \cite{FHdistmeas}.) In
this way the experimenter 'shares' the
mentioned 'knowledge' of subsystem
\$I\$ about subsystem \$II$.

(iii) In step (i) one actually arrives
at a coherent bipartite state vector
(2) with \$|\alpha|=|\beta|.\$ This
gives {\it EPR-type entanglement},
which allows distant measurement of
incompatible observables. (This is what
we call EPR-type disentanglement.) In
particular, the composite state
\$\ket{\Psi }_{I,II}\$ allows also
distant \M of the states
\$\ket{j}_{II},\enskip j=\pm\$ (cf
(12)).

(iv) Any interaction of subsystem
\$II\$ with any other system \$D_{II}\$
leads to EPR-type composite-system
states (and EPR-type entanglement
between \$I\$ and \$IID_{II}\$). One
has, e. g., the canonical Schmidt
decompositions (16) and (17), which
allow incompatible distant \m s: the
which-way one and the which-coherence

one.

(v) System \$D_{II}\$ can be the
detector for particle \$II\$ in
delayed-choice erasure, and the
interaction can be that in the
corresponding localization \m .\\

\vspace{0.5cm}

{\bf \noindent 7 Concluding Remarks}\\

\noindent {\bf A)} The assumption that
one can, in principle, {\it describe
the macroscopic detector by \qm }, like
any microscopic system, is
controversial. Let me give two facts
supporting the assumption.

(i) Niels Bohr himself felt
occasionally the need to describe
macroscopic bodies by \qm . To
illustrate this, I'll quote his words
(\cite{Bohr}, p. 50):

\begin{quote}
\indent \small ... heavy bodies like
diaphragms and shutters... . ..., in
contrast to the proper measuring
instruments, {\it these bodies}
together with the particles would
constitute the system to which the \QMl
formalism has to be applied." (Italics
by F. H.)
\end{quote}

\rm

The same was quoted by Shimony
\cite{Shimony}. He commented upon it as
follows (p. 770):

\begin{quote}
\indent \small "Bohr is saying that
from one point of view the apparatus is
described classically and from another,
mutually exclusive point of view, it is
described quantum mechanically."
\end{quote}

\rm

(ii) As it is well known, there exist
quantum systems that are macroscopic
like e. g. superconductors, quantum
fluids etc. There is a tendency to
prove experimentally that \QM is valid,
in principle, for all macroscopic
objects. Important progress was
achieved along these lines (see e. g.
\cite{meso}). The validity of \qm , in
particular, of the superposition
principle, was demonstrated, but only
for mesoscopic systems so far
(as far as known to the author).\\

\noindent {\bf B)} An argument against
treating detectors (and other
macroscopic objects) in the same way as
microscopic ones, as done in this
article, constitutes the fact that the
former are constantly and strongly
dynamically and statistically coupled
to the environment. Modern decoherence
theory \cite{Zeh1}, \cite{Zurek},
\cite{Schloss} has established this
beyond doubt. The obvious way to
circumvent the objection is to join the
environment (or at least part of it) to
the detector etc. as if it all were one
\Q system. (The point to note is that
the experimenter has to be in the
'subject' of quantum-mechanical
description.)\\

\noindent {\bf C)} Bohr's words quoted
in A(i) imply acceptance of
displacement of the so-called cut of
Heisenberg (cut between object and
observer-system or 'object' and
'subject' as it is often said). The
physics does not change in such
displacement. This has been established
in the famous psycho-physical
parallelism of von Neumann \cite{vN}
(p. 420).

Bohr's acceptance of displacement of
the cut can be viewed as {\it a smooth
connection} between the Copenhagen
interpretation \cite{Stapp} and that of
Everett \cite{Everett} if the latter is
understood as saying that the object of
\QMl description is {\it relative} to
the preparator (and, possibly, to part
of its state) and to the measuring
instrument (and, possibly, to part of
its state, primarily one displaying the
result) when one applies \QM to an
experiment. An analysis \cite{RC} of
Mott's famous quantum-mechanical
insight in the formation of
droplet-trajectories in bubble chambers
led to {\it relative-collapse} ideas (a
year earlier than Rovelli's noted
similar ideas \cite{Rovelli} that he
called 'relational').

Let me point out that von Neumann's
mentioned theorem on psycho-physical
parallelism implies that if one and the
same chain of subsystems is part of the
'object' in two different cuts, the
chain has {\it the same physical
description}. (I do not mean just
probabilities; I mean the
reality-of-state approach, which
contains the probabilities.)

To be concrete, the bipartite \$I+II\$
system in the case of simple erasure
treated in this article (an example of
the chain) is equally described in the
version when the cut is such that the
detector \$D_{II}\$ of \$II\$ is part
of the subject and in the version when
it is part of the object. This is
presented in more detail in Appendix C.

The authors of Ref. \cite{ScullyAJP99}
claim (p. 328, left column): "It
appears that Mohrhoff is led astray by
regarding the state reductions ... as
physical processes, rather than
accepting that they are nothing but
mental processes." (See more about the
controversy itself in remark H below.)

In the reality-of-state approach
adopted in this article one avoids
attributing objectivity to collapse by
utilizing a kind of Everett-like
relative state approach as stated
above. The "mental process" is then the
choice of the cut, and what belongs to
the 'object' and what to the 'subject'.
One actually chooses in this way {\it
what aspect} of the quantum reality one
wants to 'highlight', i. e., make part
of the object.

It turns out that the quantum formalism
can provide us {\it only with relative
reality}: with respect to the
preparator (and, possibly, part of its
state) and regarding our choice of the
'subject'. In a \QMl analysis of an
experiment, the experimenter is
necessarily part of the subject; he is
even the source of the decision where
the cut should be. Unfortunately, this
is far from objective, i. e.,
preparation- and
observation-independent reality, which
would have to explain also the
experimenter and his choice of the cut.

The
relative-reality-of-unitarily-evolving-state
(RRUES) approach of this article is
close to that of Dieter Zeh \cite{Zeh2}
(to mention only one of whom I am
aware). I have the impression that
the difference is mostly in emphasis.\\

\noindent {\bf D)} According to Zeh
"erasure" is a misnomer \cite{Zeh3}.
Every name given to a phenomenon must
disregard most of the aspects of the
latter, but it should catch its main
feature. Hence, perhaps "revival" or
"recoherence" \cite{Zeh2} (p. 10) would
be more appropriate. But the term
"erasure" made history, and the catchy
ring to it has a great part of the
merit for this.\\

\noindent {\bf E)} It is important to
point out that in any correlated state
vector \$\ket{\Phi}_{I,II}\$ one can
have {\it disentanglement}: measuring,
e. g., any complete observable for
subsystem \$I\$ that has nontrivial
(distant) effect on the state
\$\rho_{II}\$ of subsystem \$II\$. One
thus achieves a decomposition of
\$\rho_{II}\$. It is distant state
decomposition. A decomposition of
\$\rho_{II}\$ can be understood as a
measurement if and only if it is an
{\it orthogonal decomposition}, when we
have the special case of {\it distant
\M } \cite{FHdistmeas}.

Disentanglement produces an actually
decomposed state: the entanglement in
\$\ket{\Phi}_{I,II}\$ is eliminated.
Naturally, this is true in the
following relative sense: the cut is so
chosen that the apparatus that performs
the direct measurement on subsystem
\$I\$ is part of the 'subject'.
Contrariwise, if the apparatus is part
of the 'object', and if its initial
state is pure, the entanglement is only
elevated to the larger system. (Also
this is an example how a mixture can be
proper or improper \cite{D'Espagnat}
depending on the choice of the cut: in
the former choice one deals with actual
decomposition, i. e., with a proper
mixture; in the latter with an improper
one.)

It should be realized that every
correlated state vector
\$\ket{\Phi}_{I,II}\$ allows two
mutually incompatible distant state
decompositions of \$\rho_{II}\$. They
are performable, e. g., by measuring
two distinct complete observables for
subsystem \$I\$ that have nontrivial
distinct (distant) effects on
\$\rho_{II}\$. By definition, two state
decompositions are incompatible if they
do not allow a common further
decomposition of both. If not both
distant state decompositions are
orthogonal, i. e., \m s, one does not
have EPR-type disentanglement. But
physical interpretation may be in
conflict with quasi-classical local
thinking also in
this more general case .\\

\noindent {\bf F)} It was, of course,
quite clear to Scully et al. that
simple erasure is an EPR-type
experiment (\cite{ScullyAJP99},
Appendix 2.) It is shown in this
article that so is delayed-choice
erasure. In the second part of this
study, also the random delayed-choice
real experiment of Kim et al.
\cite{KIM} is shown to be of this
kind.\\

Englert, Scully and Walther call {\it
the EPR paradox} \cite{EPR}, addressed
at the Copenhagen claim of completeness
of quantum-mechanical description, only
"EPR problem". They also give a
solution to this problem. It is well
illustrated by the last sentence in
their paper \cite{ScullyAJP99}: "...
the property of being a first-slit atom
(particle \$II\$ in the state
\$\ket{1}_{II}\$, F. H.) is not
possessed by the atom until the photon
is found in the first resonator (until
particle \$I\$ is found in the state
\$\ket{1}_I\$, F. H.)".

The RRUES approach of this study
suggests a similar solution. Actually,
{\it there is not even an EPR problem}.
The above "property ... possessed" by
particle \$II\$ is relative to particle
\$I\$ being in the state \$\ket{1}_I\$,
or relative to the which-way detector
on which the result 'way 1' is
displayed. Thus, the seemingly local
(or subsystem) property is actually
{\it global}, and hence there is no
problem.

Let me try to explain this in case of
the original position and linear
momentum EPR-type disentanglement
\cite{EPR}. The bipartite state vector
suggested by the authors contains
precise correlation both between the
positions of the two particles and
between their linear momenta. (Let us
pretend, for the sake of argument, that
it is a normalized, i. e., a correct
state vector.) These two {\it
correlations} are compatible, and hence
both objectively present. But the very
positions and linear momenta of the
particles are only possibilities; not
actualities (in the usual sense of
these terms).

Considering collapse as a mental
process (cf second half of remark C)
implies attributing {\it reality} to
the (coherent) {\it possibilities} and
the {\it correlations} (established in
interaction of subsystems). 'Actuality'
is the product of the mentioned mental
process: when we choose the cut, then
in the object-subject relation we can
highlight part of the potentialities
elevating them to 'actualities'.

Returning to the original EPR case,
this means the following. We can take a
definite position of the first particle
in the 'subject' (of \QMl description),
then the second particle has a definite
position (in the 'object' of
description). Instead, we can measure
the position of the first particle, and
put the measurement result in the
'subject'. Then the positions of both
particles have definite values.
Analogously, we can do this for linear
momentum. What the \QMl formalism does
not allow us to do is highlighting both
position and momentum at the same time.
Nevertheless, at the potential level
they are both real.

This calls to mind Mermin's famous
mantra (in his Ithaca interpretation)
"The correlations, not the correlata"
\cite{Mermin}. I think this is \qml ly
quite true. (Seevinck's result
\cite{Seevinck}, contrary to the claim
of the author, does not seem to cast a
shadow on the reality of quantum
correlations, which are experimental
facts; instead, it seems to show that
they can be global.)

The well-known original EPR argument
\cite{EPR} assumed collapse and values
obtained in measurement as a reality.
Hence the fallacy of their conclusion
that \QM can not be considered a
complete description of an experiment.
In relative-state interpretation it
can.

The illusion of the EPR paradox will be
discussed again in connection with the
real random delayed-choice erasure
experiment of Kim et al. \cite{KIM}
near the end of the second
part of this study. \\

\noindent {\bf G)} It has turned out
that application of \QM to macroscopic
systems has interesting consequences. I
have in mind the EPR-type entanglement
described in section 4, which allows
serious distant manipulation of the
quantum state of the macroscopic body.
It appears as if the state of the
latter had kind of a classical
'surface', which is stable, robust, and
independent of influence from the
environment, and a quantum 'interior',
which can be distantly manipulated on
account of quantum correlations.\\

\noindent {\bf H)} There was an
interesting debate in the American
Journal of Physics between Mohrhoff on
the one hand and Englert, Scully and
Walther on the other. In
\cite{Mohrhoff1} Mohrhoff challenged
the idea of delayed-choice erasure, in
particular, its micromaser realization
championed by the latter authors
\cite{Nature}, \cite{ScullySciAm94}.

Taking the reality-of-states point of
view, as done also in the present
article, he pointed out that the
EPR-type correlation existing between
subsytems \$I\$ and \$II\$ in simple
erasure (cf (6a) and (9)) disappears in
the act of detection of subsystem
\$II\$ (cf the analogous remark in the
passage beneath relation (17) above).

The mentioned defenders of
delayed-choice erasure refuted
Mohrhoff's argument in two articles
\cite{ScullyFP98}, \cite{ScullyAJP99}.
They took the point of view that \QM is
no more than a book-keeping device for
calculating probabilities of future
experiments. Then, as it was already
mentioned, they showed that the
relevant probabilities are the same in
simple and delayed-choice erasures.

In a subsequent article, Mohrhoff
conceded his "error"
\cite{MohrhoffAJP99}. It is not
surprising that Mohrhoff readily gave
up the reality-of-states approach to
\QM if we remember his Pondicherry
interpretation \cite{MohrhoffAJP00} (cf
Section V there, cf also
\cite{Mohrhoff05}).\\

\noindent {\bf I)} To comment on the
apparently play-safe attitude of
Englert, Scully and Walther, I first
quote their words \cite{ScullyAJP99}
(Concluding remarks there):

\begin{quote}
\indent \small "The state vector ...
serves the sole purpose of summarizing
concisely our knowledge about the
entangled atom-and-photon system..."
\end{quote}

\rm

One cannot disagree with this unless by
"knowledge" they mean only
probabilities. That the latter is true
is suggested also by the fact that they
themselves characterize their attitude
as a "minimalistic interpretation of
state vectors ... common to all
interpretations..." (ibid. p. 328, left
column).

On the other hand, it seems to me that
the authors surpass the minimalistic
position when they say (ibid. p.327,
right column) "...this unacceptable
view (they mean that of Mohrhoff, F.
H.) results from a lack of appreciation
of the {\it objective nature} of the
EPR-type correlations that link photon
states to corresponding atom states"
(remember that photons are particles
\$I\$ and atoms are particles \$II\$ in
their micromaser case; italics by F.
H.).

Correlations are part of the
composite-system state; actually, the
former determine the latter
\cite{Mermin}. If they have an
objective nature, it is hard to imagine
how the state is deprived of it.

In support to their 'minimalistic'
point of view, the authors remind of
van Kampen's Theorem 4 \cite{Kampen}
where he says:

\begin{quote}
\indent \small "Whoever endows the
state vector with more meaning than is
needed for computing observable
phenomena is {\it responsible} for the
consequences." (Italics by F. H.)
\end{quote}

\rm

Perhaps one should take responsibility
and adopt the reality-of-states
attitude, as it is done in this
article, and thus, besides being able
to compute probabilities, also gain a
fuller insight in the \QMl 'mechanism'
of numerous beautiful thought and real
experiments. Among them stand out the
ones given to us by Scully et al. (cf
also Ref-s [5] and [6]).\\

\vspace{0.5cm}

{\bf \noindent Appendix A. On the
'partner state' in a canonical Schmidt
decomposition}\\

\noindent It was shown in previous work
that every bipartite state vector
implies an antiunitary, so-called
correlation operator \$U_a\$ that maps
the range of the first-subsystem
reduced density operator onto the
(always equally dimensional) range of
the second reduced density operator
(see a short form in subsection 2.1 in
\cite{envariance} or more details in
\cite{FHdistmeas}). The correlation
operator maps the first state vector
onto the second (partner) one in each
term of every canonical Schmidt
decomposition of the bipartite state
vector at issue.

Thus, the correlation operator comes
handy in EPR-type entanglement, where
there is a non-denumerable infinity of
distinct canonical Schmidt
decompositions (and \$U_a\$ is one
fixed entity). If one such
decomposition is given, then one can
just read \$U_a\$, and make use of it
in any other such decomposition.

In particular, we can in (6a) read that
the correlation operator implied by
\$\ket{\Psi}_{I,II}\$ is determined by
the following map of basis onto basis
(in an antilinear way):
$$U_a\ket{j}_I= \ket{j}_{II},\quad
j=1,2.\eqno{(A.1)}$$ Then (8a,b) is
immediately obtained
from (7a,b).\\

\vspace{0.5cm}

{\bf \noindent Appendix B. On simple
coherence states}\\

\noindent The requirement of symmetry
under exchange of each of the two terms
in a canonical Schmidt decomposition of
\$\ket{\Psi}_{I,II}\$, in view of
(7a,b), (8a,b) and (9) is equivalent to
$$e^{i\lambda}=e^{-i\lambda},\quad
e^{i\delta}=e^{-i\delta},\quad
e^{i\gamma}=e^{-i\gamma}.$$ This is
further equivalent to $$\lambda
=0\enskip\mbox{or}\enskip\pi,\quad
\delta
=0\enskip\mbox{or}\enskip\pi\quad
\gamma =0\enskip\mbox{or}\enskip\pi .$$

First, we take \$\gamma =0\$, because
it is anyway an open phase factor. Then
we are left with 4 possibilities, but
\$\lambda =0=\delta\$ and \$\lambda
=\pi =\delta\$ give coherence vectors
distinct by a factor \$(-1)\$, and so
do \$\lambda =0,\enskip\delta =\pi\$
and \$\lambda =\pi,\enskip\delta =0\$.
Hence, if we assume for simplicity
\$p=q\$, we are left with the two
coherence vectors in (10) as claimed.

The requirement that the exchange
operator take one term in a canonical
Schmidt decomposition into the other
is, in view of (7a,b), (8a,b) and (9),
again putting \$\gamma=0\$, equivalent
to \$p=q,\enskip\lambda =0,\enskip
e^{i(\delta +\pi )}= e^{-i\delta }\$,
and this is further equivalent to
\$\delta =\pi /2\$ or \$\delta =-\pi
/2\$ giving the two coherence vectors
of (11) as claimed.\\

\vspace{0.5cm}

{\bf \noindent Appendix C. Illustration
of the Role of the Cut in the
Relative-State Interpretation. Simple
Erasure}\\

\noindent We now take the
composite-system state
\$\ket{\Psi}_{I,II}\$ of (6a) or (12),
we introduce both detectors (or set of
detectors), \$D_I\$  for particle \$I\$
and \$D_{II}\$ for particle \$II,\$ as
part of the 'object' (of \QMl
description). Let \$M\$ denote the
choice of detector for particle \$I\$:
\$M=ww\$ in the which-way \m , and
\$M=coh\$ in the coherence \m .
Further, let zero denote the initial,
untriggered state of the detectors.
{\it At an arbitrary moment} \$t,\$
{\it the 'object' is then described by
the state vector}
$$\Big(U_{D_II}(t)U_{IID_{II}}(t)\Big)
\Big(\ket{M,0}_{D_I}\ket{\Psi}_{I,II}
\ket{0}_{D_{II}}\Big).\eqno{(C.1a)}$$

Finally, let \$d=1,2\$ be the possible
results of the \$M=ww\$ measurement,
and \$d=+,-\$ those of the coherence
measurement (cf (10)). Then, after the
choice of the experimenter whether he
wants to have which-way \M or coherence
\M (erasure), (C.1a) is suitably
rewritten in the following two
respective versions:
$$\Big(U_{D_II}(t)U_{IID_{II}}(t)\Big)
\Big[\ket{M=ww,0}_{D_I}
\Big((1/2)^{1/2}
(\ket{1}_I\ket{1}_{II}+
\ket{2}_I\ket{2}_{II})
\Big)\ket{0}_{D_{II}}\Big],\eqno{(C.1b)}$$
and
$$\Big(U_{D_II}(t)U_{IID_{II}}(t)\Big)
\Big[\ket{M=coh,0}_{D_I}
\Big((1/2)^{1/2}
(\ket{+}_I\ket{+}_{II}+\ket{-}_I
\ket{-}_{II})
\Big)\ket{0}_{D_{II}}\Big]\eqno{(C.1c)}$$
(we have substituted in (C.1a) (6a) and
(12) respectively).

Relations (C.1b) or (C.1c) describe the
process as a dynamically closed system,
i.e., when the 'object' and 'subject'
(or observing system) are dynamically
and statistically independent of each
other. (Equivalently, any enlargement
of the 'object' would include a part of
the environment state via a
non-interacting tensor product.)

In {\it simple erasure} (C.1c) is
relevant, and, at an instant \$t_I\$,
the detector \$D_I\$ is triggered
displaying either that particle \$I\$
was in the state \$\ket{+}_I\$ or that
it was in \$\ket{-}_I\$. The detection
of particle \$II\$ takes place at a
later instant \$t_{II}\$:
\$t_I<t_{II}$.

In the time interval \$t_I<t<t_{II}\$
(C.1c) becomes
$$(1/2)^{1/2}
\bigg\{\Big[U_{D_II}(t)
\Big(\ket{M=coh,d=+}_{D_I}
\ket{+}_I\Big)\Big]\Big(U_{II}(t)
\ket{+}_{II}
\Big)\Big(U_{D_{II}}\ket{0}_{D_{II}}
\Big)\quad +$$
$$\Big[U_{D_II}(t)\Big(\ket{M=coh,d=-}
_{D_I}
\ket{-}_I\Big)\Big]\Big(U_{II}(t)
\ket{-}_{II} \Big)\Big(U_{D_{II}}
\ket{0}_{D_{II}}\Big)\bigg\}.
\eqno{(C.2)}$$ (Note that the state of
photon \$II\$ and that of the detector
\$D_{II}\$ evolve independently of each
other.)

Contrariwise, {\it the simplest choice
of the cut} is such that subsystem
\$D_I\$ in the state displaying either
\$+\$ or \$-\$, particle \$I\$ and
subsystem \$D_{II} \$ are 'moved' to
the 'subject'. Then, (C.2) is replaced
by \$\Big(U_{II}\ket{+}_{II}\Big)\$ or
by \$\Big(U_{II}\ket{-}_{II}\Big)\$
respectively. (This is the most common
\QMl 'picture' of simple erasure.)

One can displace the cut half a step
back: leave the same subsystems in the
'subject', but without specifying the
result of \M on \$D_I\$. Then instead
of (C.2) we have the density operator
$$\rho_{II}=(1/2)\Big(U_{II}
\ket{+}_{II}\bra{+}_{II}U_{II}^{\dag}
\quad +\quad
U_{II}\ket{-}_{II}\bra{-}_{II}
U_{II}^{\dag}\Big). \eqno{(C.3)}$$

Note that also (C.1c) implies
\$\rho_{II}\$ of (C.3), but as an
improper mixed state (cf
\cite{D'Espagnat}). In case of the
displaced cut defining (C.3),
\$\rho_{II}\$ is a proper mixture,
expressing the lack of knowledge on
part of the 'subject' what the result
of the measurement by \$D_I\$ is.\\

\pagebreak

{\bf \noindent Appendix D. The Equality
of Probabilities in Simple and
Delayed-choice Erasures}\\

{\noindent \bf Simple erasure}\\

\noindent Detector \$D_I\$ detects one
of the four possibilities
(\$M=ww,\enskip d=1,2;\enskip M=coh,
\enskip d=\pm\$) for particle \$I\$
(see the preceding Appendix for more
details). Detector \$D_{II}\$ is that
of localization of particle \$II\$. One
can imagine that \$D_{II}\$ consists of
a number, say \$N\$, detectors
\$\{D_{II}^n:n=1,2\dots ,N\}\$ placed
along the vertical \$x\$ axis. Let us
denote by \$\ket{d=1}_{II}\$ or
\$\ket{d=2}_{II}\$ the state vector of
particle \$II\$ if one has distant
which-way \m , and we denote it by
\$\ket{d=+}_{II}\$ or
\$\ket{d=-}_{II}\$ if the distant \M is
of the coherence type. Then immediately
before detection, occurring at
\$t_{II}\$, we have the decomposition
$$\ket{d,t_{II}-\epsilon}_{II}=
\sum_{n=1}^N \int_{x_n-(\Delta
x)/2}^{x_n+(\Delta
x)/2}\Big(\bra{x}_{II}\ket{d,t_{II}-
\epsilon}_{II}\Big)\ket{x}_{II}dx,\quad
0<\epsilon\ll 1,\eqno{(D.1a)}$$ where
\$x_n\$ is at the center of the \$n$-th
detector, \$\Delta x\$ is the width of
the detectors, and \$N\Delta x\$ is a
large enough span so that the photon is
with certainty detected (in one of the
\$N\$ detectors):
$$\Big|\int_{x_1-(\Delta
x)/2}^{x_N+(\Delta
x)/2}\Big(\bra{x}_{II}\ket{d,t_{II}
-\epsilon}_{II} \Big)dx\Big|^2 =1.
\eqno{(D.1b)}$$

State-vector decomposition (D.1a) can
be viewed as a {\it coherent} set of
\$N\$ {\it possibilities} (expressed by
the \$N\$ component terms in (D.1a)),
of which precisely one will be realized
in the localization measurement for the
individual particle \$II\$. The
probability \$p_n(d)\$ of the \$n$-th
possibility to be realized is, as it is
well known,
$$p_n(d)=\lim_{\epsilon\rightarrow 0}
\Big|\int_{x_n-(\Delta
x)/2}^{x_n+(\Delta x)/2}
\bra{x}_{II}\ket{d,t_{II}
-\epsilon}_{II}\Big)dx \Big|^2
\eqno{(D.2)}$$ because coherence
implies adding (this time via
integrals) the amplitudes and not the
probabilities.

Note that the certain detection in
\$D_{II}\$ expressed by (D.1b) has to
be replaced by \$\sum_{n=1}^Np_n(d)=1\$
in spite of the coherence in (D.1a).
Namely, when the measurement is
performed, it deletes the coherence
expressed by the sum in (D.1a).

Since \$\ket{d,t_{II}-\epsilon}_{II}=
U_{II}(t_{II}-\epsilon)\ket{d}_{II}\$,
substitution in (C.2) makes it clear
that the {\it probability} of the
result \$d=\pm\$ of distant coherence
measurement and of localization in
detector \$D_n\$ for simple erasure is
$$p_{coh}^{simple}(d,n)=(1/2)p_n(d).
\eqno{(D.3)}$$

In case of which-way distant \m , the
probability formula is analogous.\\

{\noindent \bf Delayed-choice erasure}\\

\noindent This erasure is characterized
by \$M=coh\$, \$d=\pm\$, and
\$t_{II}<t_I\$, i. e., localization of
particle \$II\$ is detected earlier
than the detection of particle \$I$
takes place.

The state vector (C.1c) then becomes in
the time interval \$t_{II}<t\leq t_I\$:
$$(1/2)^{1/2}\bigg\{\Big[
\Big(U_{D_I}(t)
\ket{M=coh,0}_{D_I}\Big)\Big(U_I(t)
\ket{+}_I\Big)\Big]\otimes
\Big[U_{IID_{II}}(t)\Big(
\ket{+}_{II}\ket{0}_{D_{II}}\Big)\Big]
\quad +$$
$$\Big[\Big(U_{D_I}(t)
\ket{M=coh,0}_{D_I}\Big)
\Big(U_I(t)\ket{-}_I\Big)\Big]\otimes
\Big[U_{IID_{II}}(t)\Big(
\ket{-}_{II}\ket{0}_{D_{II}}\Big)\Big]
\bigg\}.\eqno{(D.4)}$$ (Note that this
time the state of detector \$D_I\$ and
that of photon \$I\$ evolve
independently of each other.)

It is suitable to replace \$t\$ by
\$t'\equiv t-t_{II}\$, i. e., to
consider the time from the detection of
particle \$II\$ (and not from leaving
the slits). For simplicity, we write
again \$t\$ instead of \$t'\$.

Taking the limiting form of relation
(D.1a) for \$\epsilon\rightarrow 0\$,
for the relevant results \$d=\pm\$, and
omitting \$t=0\$, one has
$$\ket{d=\pm}_{II}= \sum_{n=1}^N
\int_{x_n-(\Delta x)/2}^{x_n+(\Delta
x)/2}\Big(\bra{x}_{II}\ket{d=\pm}_{II}\Big)
\ket{x}_{II}dx.$$

Substituting this in (D.4), one obtains
$$(1/2)^{1/2}\sum_{n=1}^N\sum_{d=\pm}
\Big\{\Big[\Big(U_{D_I}(t)
\ket{M=coh,0}_{D_I}\Big)\Big(U_I(t)
\ket{d=\pm}_I\Big]\otimes$$
$$\Big[U_{IID_{II}}(t)
\Big( \int_{x_n-(\Delta
x)/2}^{x_n+(\Delta
x)/2}\Big(\bra{x}_{II}\ket{d=\pm}_{II}\Big)
\ket{x}_{II}dx\ket{0}_{D_{II}}\Big)
\Big]\Big\}.\eqno{(D.5)}$$

Now we can ask the question what is the
probability \$p_{coh}^{del.-ch.}(d,n)\$
of choosing to do coherence distant \m
, i. e., of having \$d=\pm\$ for
particle \$II\$, and of its detection
in the detector \$D_{II}^n\$ at
\$t_{II}\$. The composite-system state
vector in (D.5) is suitably decomposed,
and we can read the answer. It is
$$p_{coh}^{del.-ch.}(d,n)=
(1/2)\Big |\int_{x_n-(\Delta
x)/2}^{x_n+(\Delta
x)/2}\Big(\bra{x}_{II}
\ket{d=\pm}_{II}\Big)
\ket{x}_{II}dx\Big |^2.\eqno{(D.6)}$$

In view of (D.3) and (D.2), (D.6)
implies the claimed result
$$p_{coh}^{del.-ch.}
(d=\pm,n)=p_{coh}^{simple}(d=\pm,n)
\quad n=1,2,\dots ,N.
\eqno{(D.7)}$$\\

\end{document}